# α-FeAs-free SmFeAsO$_{1-x}$F$_x$ by Low Temperature Sintering with Slow Cooling

Masaya Fujioka[1]*, Saleem J. Denholme [1], Toshinori Ozaki[1], Hiroyuki Okazaki[1], Keita Deguchi[1,2], Satoshi Demura[1,2], Hiroshi Hara[1,2], Tohru Watanabe[1,2], Hiroyuki Takeya[1], Takahide Yamaguchi[1], Hiroaki Kumakura[1], and Yoshihiko Takano[1,2]

[1] *National Institute for Materials Science, 1-2-1 Sengen, Tsukuba, Ibaraki 305-0047, Japan*
[2] *University of Tsukuba, 1-1-1 Tennodai, Tsukuba, Ibaraki 305-0001, Japan*

We obtained amorphous-FeAs-free SmFeAsO$_{1-x}$F$_x$ using a low temperature sintering with slow cooling. SmFeAsO$_{1-x}$F$_x$ is sintered at 980 °C for 40 hours and cooled slowly down to 600 °C. The low temperature sintering suppresses the formation of amorphous FeAs, and the slow cooling introduces much fluorine into SmFeAsO$_{1-x}$F$_x$. The superconductivity of this sample appears at 57.8 K and the superconducting volume fraction reaches 96 %. To study the change of fluorine concentration during the cooling process, samples are quenched by water at 950 °C, 900 °C, 850 °C, 800 °C, 750 °C and 700 °C. It is found that fluorine is substituted not only at the maximum heating temperature but also during the cooling process. The low temperature sintering with slow cooling is very effective to obtain a homogeneous SmFeAsO$_{1-x}$F$_x$ with high fluorine concentration.

*E-mail address: FUJIOKA.Masaya@nims.go.jp

## 1. Introduction

Since the discovery of superconductivity in LaFeAsO$_{1-x}$F$_x$ (La-1111),[1] many kinds of iron-based superconductors have been reported.[2-5] Iron-based superconductors have a high upper critical magnetic field ($H_{c2}$). This advantage could make them suitable for several applications.[6-8] Of all of the iron based superconductors, SmFeAsO$_{1-x}$F$_x$ (Sm-1111) shows the highest superconducting transition temperature ($T_c$) and $H_{c2}$.[9] This material was firstly synthesized by Chen *et al* with a $T_c$ of only 43 K.[2] By optimizing the fluorine concentration and a high temperature sintering of about 1200 - 1350 °C, the $T_c$ quickly increase to 52-55 K.[10,11] The obtained Sm-1111 made by the high temperature sintering contains amorphous FeAs.[12-16] It complexly co-exists with the superconducting grains in the sample and seems to cover these superconducting grains. Therefore, amorphous FeAs possibly prevents a superconducting current from flowing between the superconducting grains.[15,16]

In our previous study, it is found that indium addition to the polycrystalline SmFeAsO$_{1-x}$F$_x$ removes the amorphous FeAs and induces the clustering of the superconducting grains. As a result, magnetic $J_c$ is increased because of an increase in the total contact surface of the superconducting grains.[17] Therefore, it is important to explore the synthesis method to prevent the formation of amorphous FeAs. On the other hand, Wang *et al* synthesized Sm-1111 by a low temperature sintering. When the nominal amount of fluorine is x = 0.2 and the heating temperature is 1000 °C, the obtained $T_c$ is 56.1 K.[18] In this study, the higher $T_c$ is observed by the low temperature sintering rather than the high temperature sintering. However, there are no reports mentioning the existence of the amorphous FeAs between the high and low temperature sintering.

The most difficult problem for the synthesis of a 1111 system, is how to control the fluorine concentration. Fluorine included in the Sm-1111 is very sensitive to heat treatment.[8] Even if Sm-1111 had been synthesized at high temperature e.g. 1200 °C, it partly decomposes into SmOF and other impurities even when re-sintering at 700 °C. At the present time, the behavior of fluorine during the heat treatment has not been still established. Therefore, the study on the change of fluorine concentration at each temperature during heat treatment is required to obtain Sm-1111 with high fluorine concentration.

In this study, we successfully synthesized amorphous-FeAs-free Sm-1111 by using low temperature sintering and discovered that the slow cooling is very effective to substitute much fluorine.

## 2. Experimental Procedure

For the preparation of SmFeAsO$_{1-x}$F$_x$, stoichiometric Sm$_2$O$_3$, SmF$_3$, and two precursors, which are named 133 and 233 powder,[8,17] were ground in a mortar and compressed into pellets. The precursors were composed of compounds made of Sm , Fe and As. The obtained pellets were heated at 1200 °C or 980 °C for 40 h in an evacuated quartz tube. After then, from a maximum heating temperature, samples were furnace cooled or slowly cooled at the rate of -5 °C/h or -3 °C/h down to 600 °C and followed by furnace cooling.

To examine how to change the fluorine concentration in the sample sintered at 980 °C during the heat treatment, the samples during slow cooling at a rate of -5 °C/h were taken out one by one from an electronic furnace and quenched by water at the following temperatures (950 °C, 900 °C, 850 °C, 800 °C, 750 °C, 700 °C) as shown in the inset of figure 4. These samples are listed in table 1. The cooling process, $T_c^{onset}$, $T_c^{zero}$ and the cell volume for each sample are described in table 1.

The obtained samples were characterized by X-ray diffraction (XRD; Rigaku Rint 2500) using Cu Kα radiation. In this study, the amount of fluorine in the $SmFeAsO_{1-x}F_x$ is estimated by the change in cell volume. To exactly calculate the cell volume, Si powder (RIGAKU: RSRP-43275G) was used as an internal standard material. Electrical resistivities of samples were measured by the standard four-probe technique using Au electrodes at the temperatures between 45 and 65 K. In this study, $T_c^{onset}$ is regarded as the cross point of the fitting lines for resistivity in the normal state near transition and in the drop area during the transition. $T_c^{zero}$ is also regarded as the cross point of the line for zero resistivity and the $\rho$-$T$ curve. Magnetic measurements were performed for the sample heated at 980 °C with slow cooling at a rate of -3 °C/h down to 600°C by SQUID magnetometer (Quantum Design: MPMS). Zero-field cooling (ZFC) and a field cooling (FC) were measured under 1 Oe. To discuss the homogeneity of the polycrystalline $SmFeAsO_{1-x}F_x$, the polished surfaces of $SmFeAsO_{1-x}F_x$ were observed by the scanning electron microscopy (SEM; HITACHI SU-70).

## 3. Results and discussion

### 3.1 The effect of low temperature sintering with slow cooling

Figure 1 shows the SEM image of the polished surface of $SmFeAsO_{1-x}F_x$. A cooling rate of the samples in the figure is -5 °C/h. (a) and (b) denote $SmFeAsO_{0.9}F_{0.1}$ sintered at 980 °C with slow cooling and furnace cooling. Impurity phases are hardly observed in both samples. When the fluorine concentration increases up to x = 0.2, it is clear that the furnace cooled sample includes SmAs impurity phases as shown in figure 1 (d). However, the impurity phases are suppressed in the slow cooled sample in spite of high fluorine substitution (x = 0.2). Whereas, in figure 1 (e) and (f), amorphous FeAs is noticeable. This impurity is inevitably produced by using high temperature sintering. Particularly in the case of figure 1 (e), much larger FeAs areas are observed by using slow cooling. This is because the sample was heated above 1000°C for a long time. We found the sintering below 1000 °C dose not form amorphous FeAs impurity phase.

Figure 2 shows XRD patterns of $SmFeAsO_{0.8}F_{0.2}$ sintered at 980 °C with furnace cooling and slow cooling at a rate of -3 and -5 °C/h down to 600°C. Black bars at the bottom show the calculated Bragg diffraction positions of $SmFeAsO_{1-x}F_x$. They correspond to the obtained XRD peaks of samples. Weak impurity peaks are also detected. Inset shows an impurity peak of SmOF at around 2θ = 28 ° The SmOF impurity decreases with slow cooling.

Figure 3 shows the temperature dependence of resistivity for $SmFeAsO_{0.8}F_{0.2}$ sintered at 980 °C with furnace cooling and slow cooling at a rate of -3 and -5 °C/h down to 600°C. It is found that slow cooling is very effective to obtain the high superconducting transition temperature. A slower cooling rate is thought to be better for the superconductivity. The sample cooled at the rate of -3 °C/h shows a higher transition temperature than at a rate of -5 °C/h. The superconducting transition temperature increases up to 57.8 K for the slow cooling. In all samples, the behavior of resistivity gradually approaches zero. However, $T_c^{zero}$ is also improved for the slower cooling. It achieved a $T_c^{zero}$ of 54.3 K

The sample cooled at a rate of -3 °C/h shows the highest superconducting transition in this study. The magnetic measurements are performed for this sample. Figure 4 shows the temperature dependence of magnetic susceptibility. Superconducting shielding volume fraction is 96 % at 5 K. The $T_c$ is obtained at 57.4 K, and FC and ZFC separate at around 51.5 K. As shown in the inset of figure 4, $M$-$T$ curve obviously shows the two distinct steps. The step at the higher temperature is due to the intragrain shielding. Whereas the lower one is due to the intergrain shielding. This behavior has been previously reported [19, 20]. Figure 5 shows the magnetic $J_c$ versus magnetic field and the inset shows the $M$-$H$ curve at 5 K. The magnetic $J_c$ is calculated through the following equation: $J_c = 20 \Delta M / a (1-a/3b)$, where $\Delta M$ is the hysteresis loop width, and $a$ and $b$ are the dimensions of the rectangular sample ($a < b$). The magnetic $J_c$ obtained a maximum value of over 10000 A/cm$^2$ around 0.25 T and gradually decreases with increasing magnetic field. This value is higher than that of the sample sintered at low temperature shown in the previous report [21]. This improvement is thought to be due to the slow cooling effect as a rate of -3 °C/h.

From these results, high temperature sintering produce the FeAs impurity phase even if the fluorine concentration is x = 0.1. On the other hand, the low temperature sintering does not form any impurity phases, when x = 0.1. However, when x = 0.2, the sample with furnace cooling includes large SmAs impurity areas. These impurity areas are suppressed by using slow cooling. It is found that low temperature sintering with slow cooling is very effective to obtain the homogeneous SmFeAsO$_{1-x}$F$_x$ with high fluorine concentration and various superconducting properties are improved for this synthesis method.

*3.2 The change in a fluorine concentration during slow cooling*

Figure 6 shows the cell volume versus the temperature at which the samples were quenched during the slow cooling process. Inserted figure shows the reaction scheme. The orange solid circle indicates the sample without quenching. Cell volume decreases with a decrease in quenching temperature. Non-quenched sample shows the smallest cell volume. Moreover, fluorine concentration is estimated from the relationship between the cell volume and fluorine concentration in previous research. [10)] Between the quenched samples at 950 °C and non-quenched sample, a large difference in fluorine concentration of about 2.41 % is observed. It is found that substitution for fluorine is performed not only at the maximum heating temperature but also during the cooling process. $T_c^{onset}$ and $T_c^{zero}$ are listed in the table 1.

Figure 7 shows the temperature dependence of resistivity for samples with each quenching temperature. Upper inset shows the expanded view near $T_c^{onset}$. Increase in $T_c^{onset}$ is almost saturated for the samples quenched below 850 °C. Lower inset shows the expanded view near $T_c^{zero}$. Colored fitting curves are calculated by a polynomial approximation. $T_c^{zero}$ increases with decreasing quenching temperature. This means that $T_c^{zero}$ is gradually closer to $T_c^{onset}$ with slow cooling down to a lower temperature. $T_c^{zero}$ is decided by the weak area of superconductivity. Therefore, this behavior is thought to be due to an improvement in these weak-superconducting areas caused by the scarce fluorine concentration. This also suggests that high fluorine concentration seems to be more widely spread in the sample as a whole during the slow cooling.

We found that the slow cooling introduces much fluorine into Sm-1111 and enhanced the superconductivity up to 57.8 K when the cooling rate is -3 °C/h. We could hope to obtain much higher $T_c$, because the fluorine concentration has not reached an optimal level.

## 4. Conclusions

We discovered an effective way to obtain much fluorine doped $SmFeAsO_{1-x}F_x$ without impurities. Whereas, a high temperature sintering of 1200 °C produces an amorphous FeAs impurity phase, a low temperature sintering of 980 °C does not form that impurity. However, when increasing the fluorine concentration, a SmAs impurity phase forms in $SmFeAsO_{1-x}F_x$, even if low temperature sintering is used. Slow cooling suppresses the formation of this impurity phase. Moreover, it is found that fluorine is substituted not only at the maximum heating temperature but also during the cooling process. A fluorine concentration of 2.41 % was introduced during the slow cooling process. To obtain much fluorine doped sample, slow cooling is very effective. We obtained the highest superconducting transition at 57.8 K in this study and the superconducting volume fraction achieves 96% by using low temperature sintering with slow cooling.

**Acknowledgments**

This work was supported in part by the Japan Society for the Promotion of Science through the 'Funding Program for World-Leading Innovative R&D on Science Technology (FIRST)' and Japan Science and Technology Agency through the Strategic International Collaborative Research Program (SICORP-EU-Japan).

Fig. 1
SEM images of the polished surface of samples. (a): SmFeAsO$_{0.9}$F$_{0.1}$ sintered at 980 °C with the slow cooling. (b): SmFeAsO$_{0.9}$F$_{0.1}$ sintered at 980 °C with the furnace cooling. (c): SmFeAsO$_{0.8}$F$_{0.2}$ sintered at 980 °C with the slow cooling. (d): SmFeAsO$_{0.8}$F$_{0.2}$ sintered at 980 °C with the furnace cooling. (e): SmFeAsO$_{0.9}$F$_{0.1}$ sintered at 1200 °C with the slow cooling. (f): SmFeAsO$_{0.9}$F$_{0.1}$ sintered at 1200 °C with the furnace cooling. A slow cooling rate is -5 °C/h in these samples.

Fig. 2
(Color online) XRD patterns of SmFeAsO$_{0.8}$F$_{0.2}$ sintered at 980 °C. Black line denotes the sample furnace cooled down to room temperature. Blue and red lines denote the sample slow cooled at a rate of -5 and -3 °C/h down to 600 °C respectively and then furnace cooled down to room temperature. Bottom bars indicate Bragg diffraction positions for SmFeAs(O,F). Inset shows the impurity peak of SmOF for each sample.

Fig. 3
(Color online) Resistivity versus temperature for SmFeAsO$_{0.8}$F$_{0.2}$ sintered at 980 °C. Black circle denotes the sample furnace cooled down to room temperature. Blue and red circles denote the sample slow cooled at a rate of -5 and -3 °C/h down to 600 °C respectively and then furnace cooled down to room temperature. Inset shows the expanded view near $T_c^{onset}$. Black lines are fitted lines for an estimation of $T_c^{onset}$.

Fig. 4
(Color online) Magnetic susceptibility versus temperature for the samples sintered at 980 °C with the slow cooling at a rate of -3 °C/h. Inset shows the expanded view near $T_c^{onset}$ of the samples. Magnetic field is 1 Oe.

Fig. 5
(Color online) Magnetic $J_c$ versus magnetic field for the samples sintered at 980 °C with the slow cooling at a rate of -3 °C. Inset shows the hysteresis loop.

Fig. 6
(Color online) Cell volume versus quenching temperature. Inset shows a heating condition. Colored points denote the quenching temperature of each sample. These colors are corresponding with the color plots of the cell volume.

Fig. 7

(Color online) Normalized resistivity at 65 K versus temperature. These samples are taken out from furnace at each quenching temperature. Upper inset shows the expanded view near $T_c^{onset}$. Lower inset shows the expanded view near $T_c^{szero}$. Colored fitting curves in the insert are calculated by polynomial approximation.

Table 1. $T_c$ and cell volume for the samples sintered at 980 °C for 40 hours with different cooling processes. The following arrows denote the difference in the cooling method. ( ⇒: furnace cooling, →: slow cooling (-5 °C/h), ↓ : quenching )

| Cooling process | $T_c^{\text{onset}}$ (K) | $T_c^{\text{zero}}$ (K) | Cell Volume (nm³) |
|---|---|---|---|
| 980 °C ⇒ room temp. | 56.9 | 51.8 | 130.716 |
| 980 °C → 600 °C ⇒ room temp. | 57.6 | 53.4 | 130.567 |
| 980 °C → 700 °C ↓ room temp. | 57.6 | 52.9 | 130.610 |
| 980 °C → 750 °C ↓ room temp. | 57.6 | 52.6 | 130.623 |
| 980 °C → 800 °C ↓ room temp. | 57.6 | 52.3 | 130.616 |
| 980 °C → 850 °C ↓ room temp. | 57.6 | 51.7 | 130.700 |
| 980 °C → 900 °C ↓ room temp. | 57.5 | 48.4 | 130.748 |
| 980 °C → 950 °C ↓ room temp. | 57.0 | 47.2 | 130.794 |

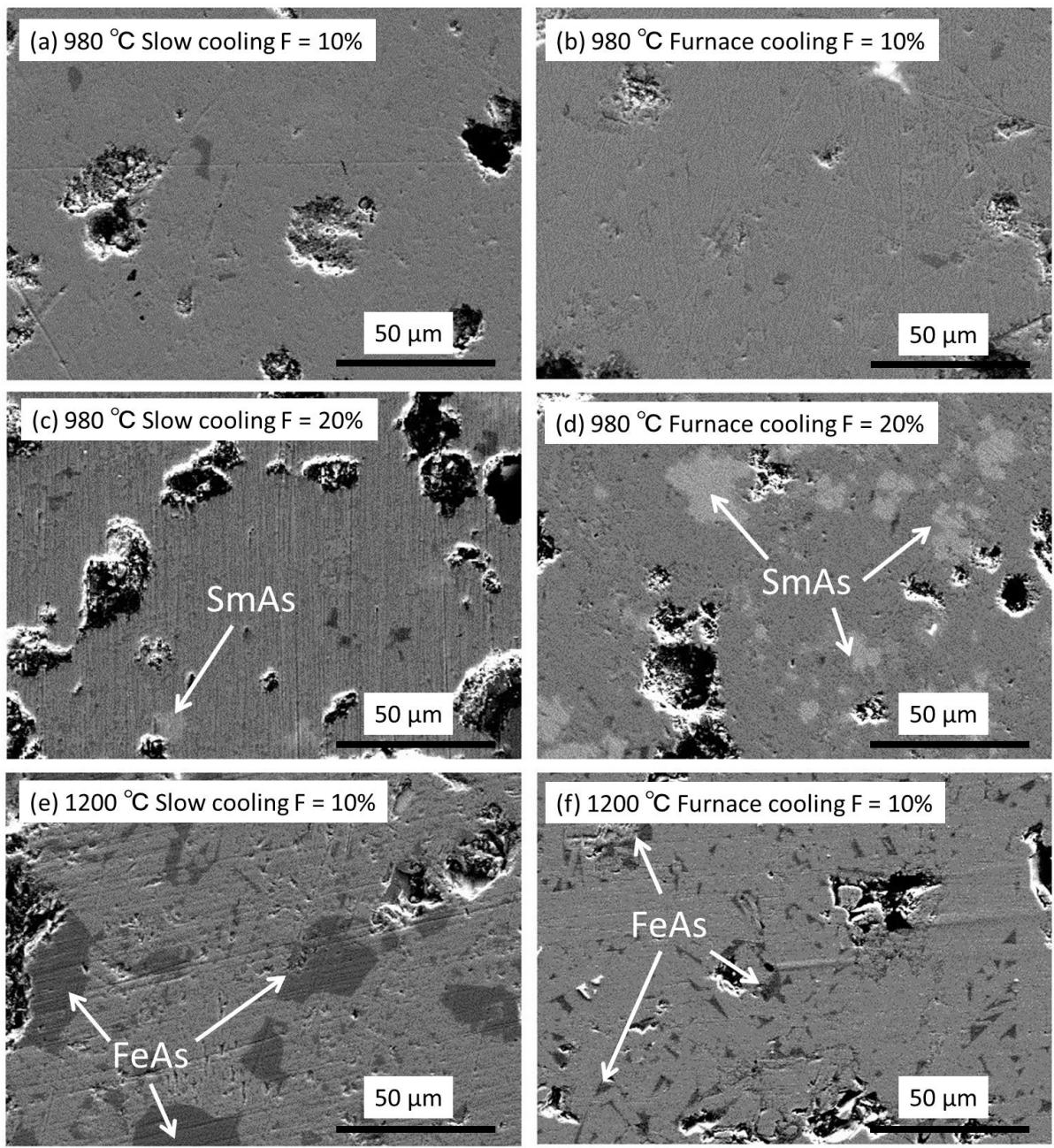

Fig. 1

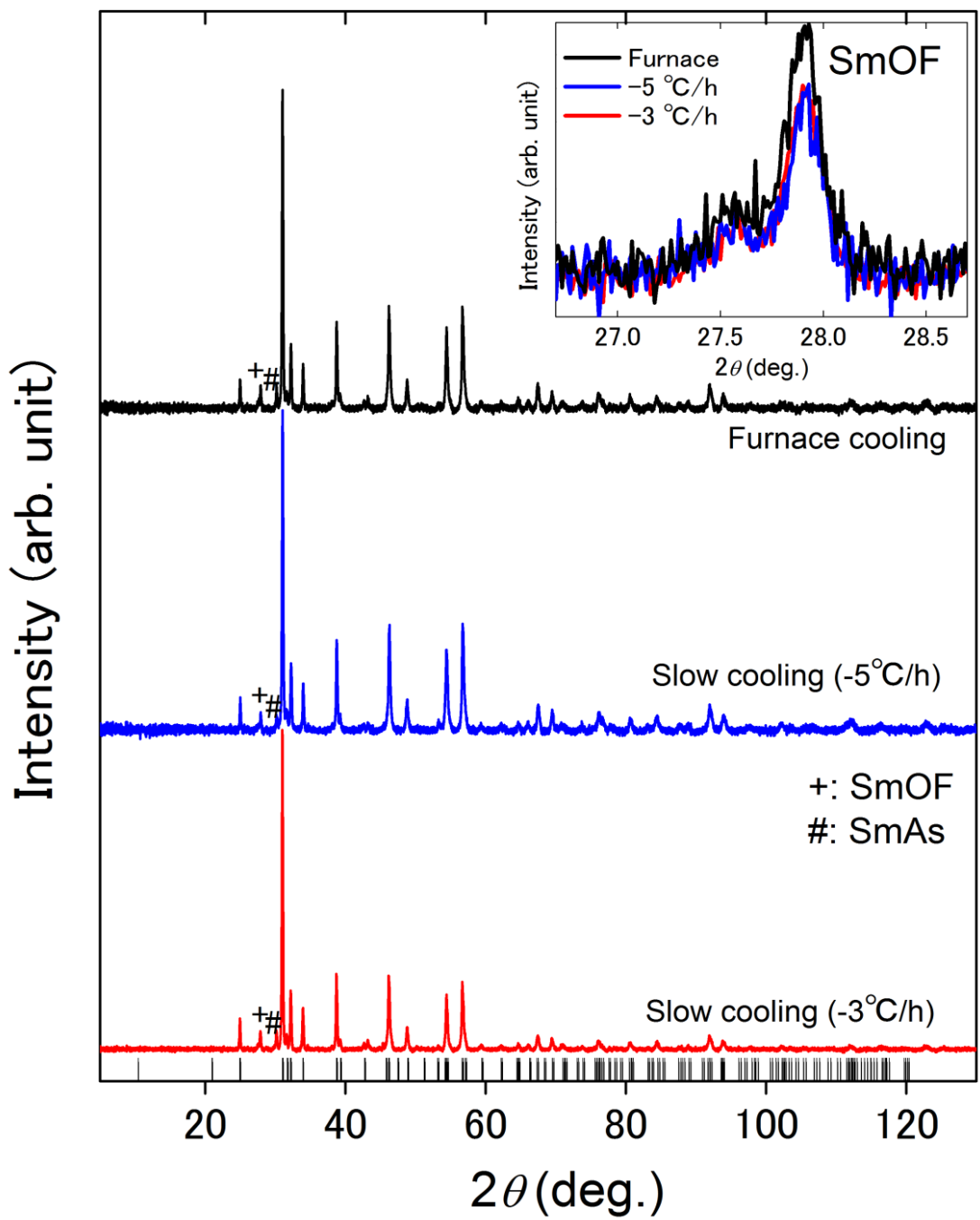

Fig. 2

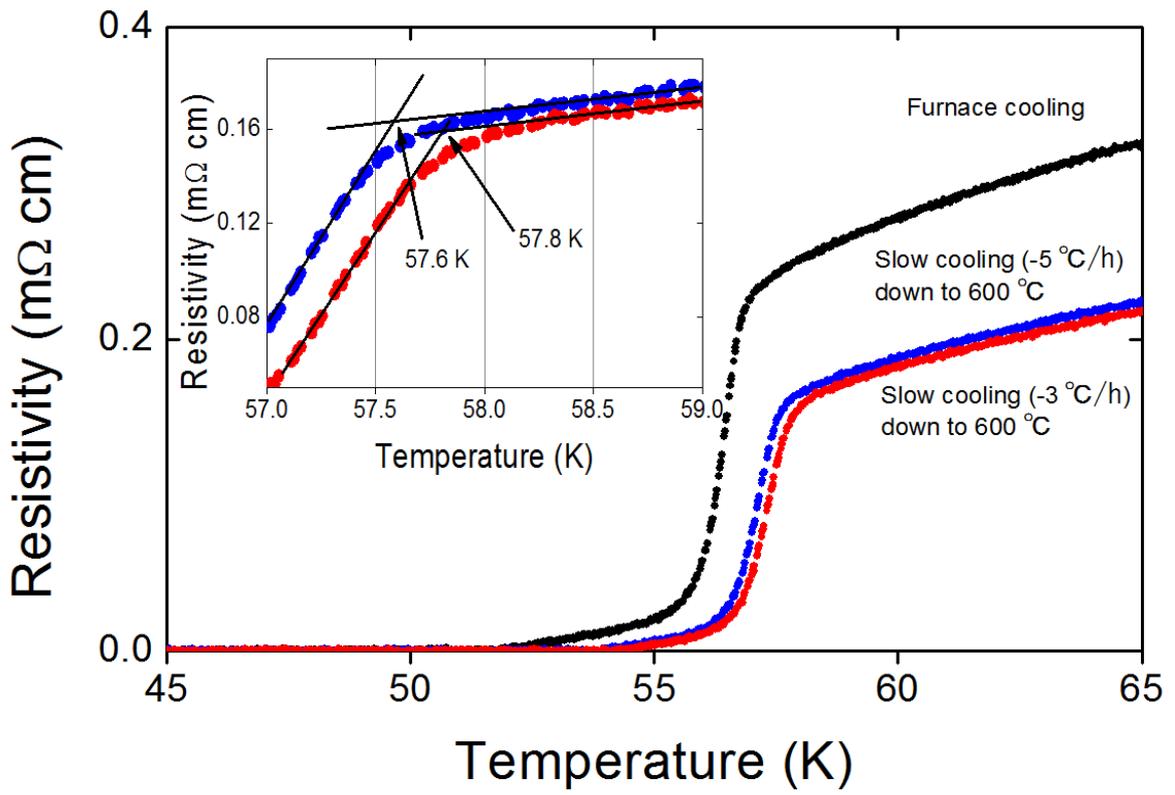

Fig. 3

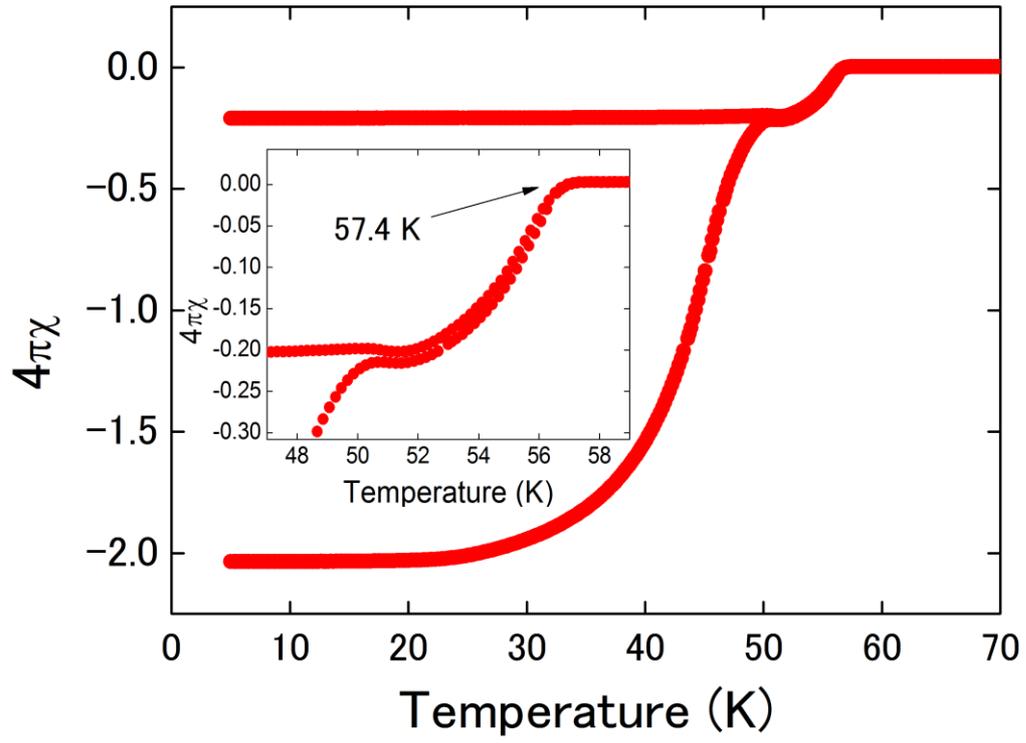

Fig. 4

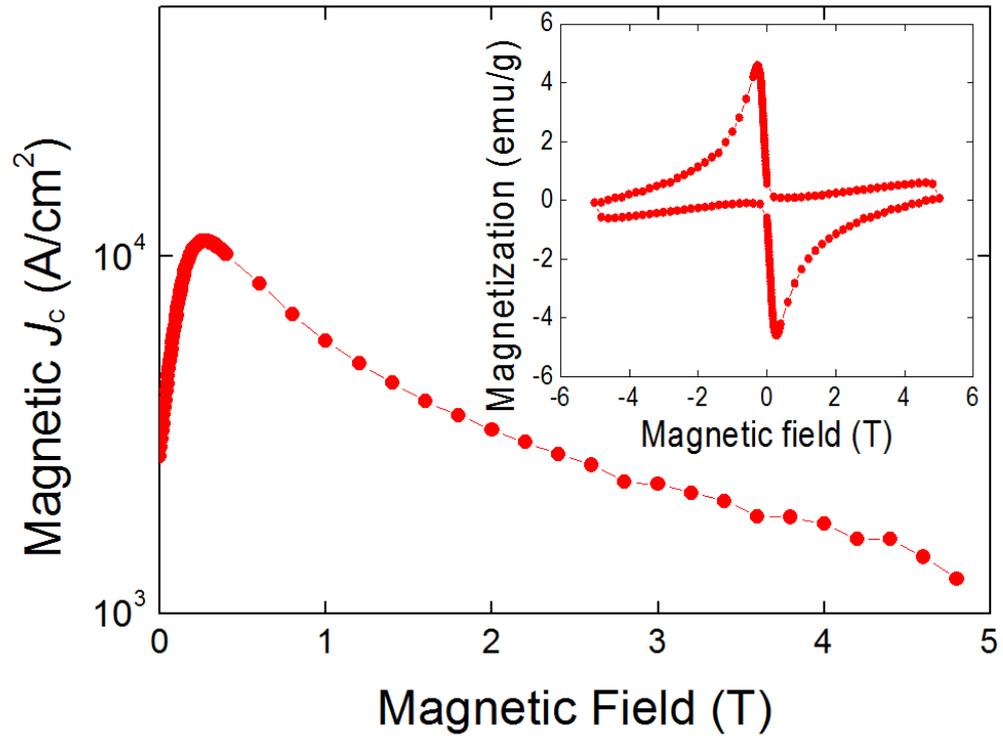

Fig. 5

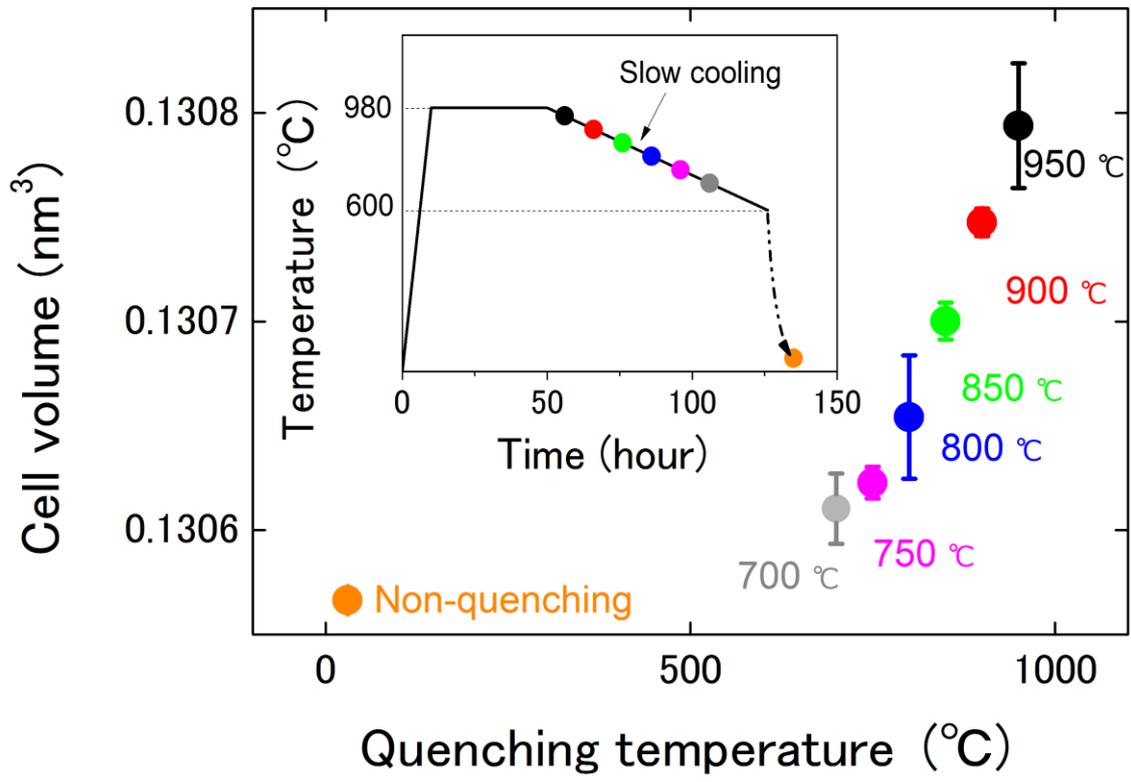

Fig. 6

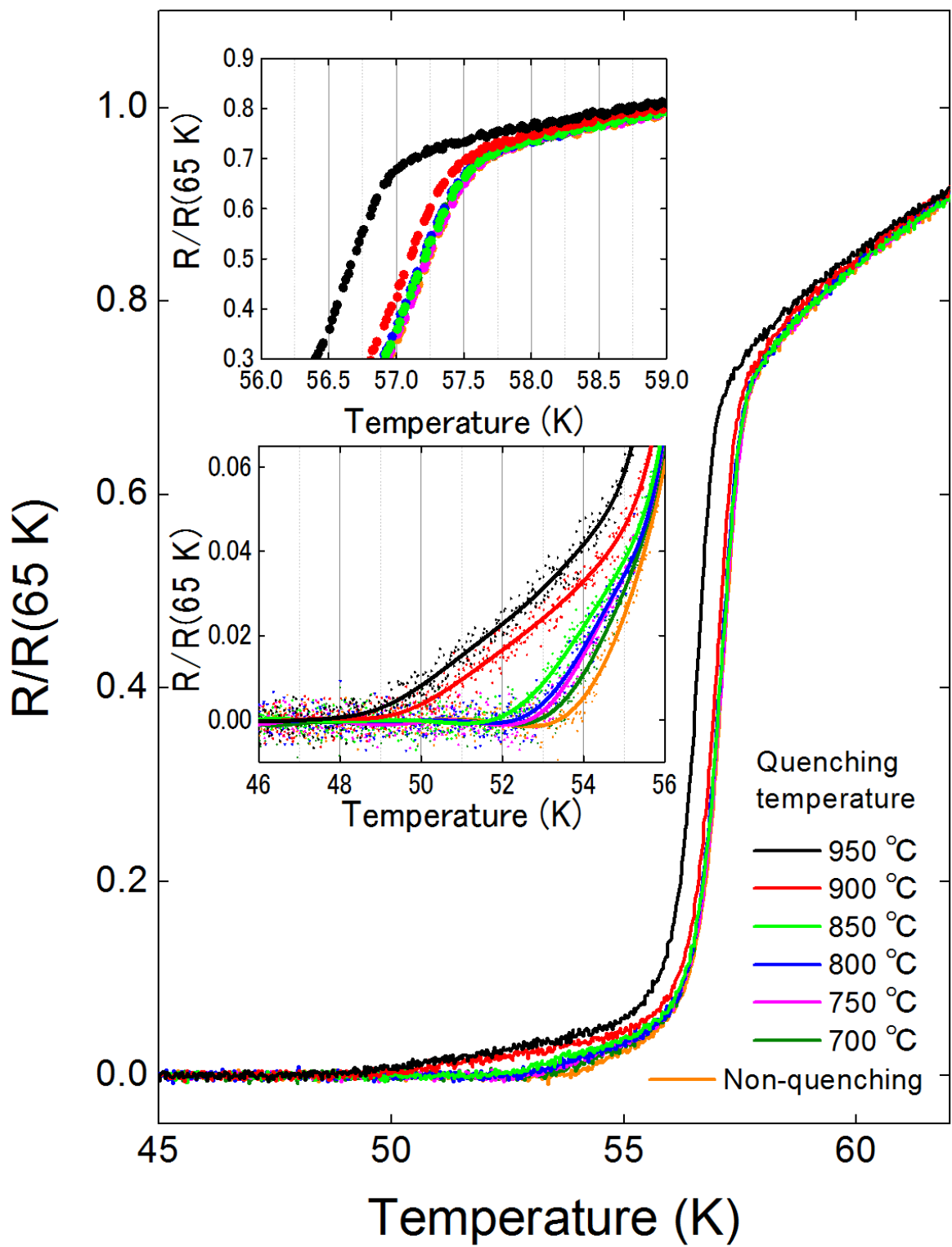

Fig. 7